%% file: main.tex
\documentclass[sigconf]{acmart}
\setcopyright{none}
\makeatletter
\let\footnotetextcopyrightpermission\@gobble
\makeatother
\usepackage{amsmath}
\usepackage{algorithmic}
\usepackage{graphicx}
\usepackage{textcomp}
\usepackage{xcolor}
\usepackage[normalem]{ulem}
\usepackage{hyperref}
\usepackage{stmaryrd}
\usepackage{aligned-overset}

\usepackage{subcaption}
\usepackage{multirow}
\usepackage{wrapfig}
\usepackage{enumitem}
\usepackage{pifont}
\DeclareRobustCommand{\cmark}{\textcolor{green}{\ding{51}}}
\DeclareRobustCommand{\xmark}{\textcolor{red}{\ding{55}}}
\DeclareRobustCommand{\pmark}{\textcolor{blue}{%
  \ensuremath{\pmb{\checkmark}}
  \kern-1.1ex
  \raisebox{.7ex}{\textbf{\rotatebox[origin=c]{125}{--}}}%
}}

\newcommand{\myparagraph}[1]{%
  \noindent\textbf{#1.}
}

\newcommand{\Toolg}{\texttt{TAC-GB}}

\begin{document}

\title{TAC: Hybrid IAM Privilege Escalation Detection}

\author{Yang Hu}
\affiliation{%
  \institution{The University of Texas at Austin}
  \city{Austin}
  \state{Texas}
  \country{USA}
}
\email{huyang@utexas.edu}
\authornote{These authors contributed equally to this work.}

\author{Wenxi Wang}
\affiliation{%
  \institution{The University of Virginia}
  \city{Charlottesville}
  \state{Virginia}
  \country{USA}
}
\email{wenxiw@virginia.edu}
\authornotemark[1]

\author{Sarfraz Khurshid}
\affiliation{%
  \institution{The University of Texas at Austin}
  \city{Austin}
  \state{Texas}
  \country{USA}
}
\email{khurshid@ece.utexas.edu}

\author{Mohit Tiwari}
\affiliation{%
  \institution{The University of Texas at Austin}
  \city{Austin}
  \state{Texas}
  \country{USA}
}
\email{tiwari@austin.utexas.edu}

\input{abstract}

\maketitle

\input{intro}
\input{background}
\input{motivation}
\input{whitebox}
\input{greybox}
\input{dataset}
\input{eval}

\input{experiments}
\input{discussion}
\input{related_work}
\input{conclusion}
\bibliographystyle{ACM-Reference-Format}
\bibliography{myref}

\end{document}

%% file: abstract.tex
\begin{abstract}
Identity and Access Management (IAM) misconfigurations are a primary cause of privilege escalation (PE) attacks on cloud platforms, leading to data breaches and significant economic losses. Yet existing PE detectors suffer from two limitations: each covers only a subset of PE types, leaving many escalations undetected; and all require full configuration access, a significant barrier when customers are unwilling to disclose sensitive organizational details.

We present TAC, the first hybrid IAM PE detection framework supporting both whitebox and greybox analysis, with Amazon Web Services (AWS) as the target platform. To address the first limitation (limited PE coverage), we conduct a systematic study of permission acquisition in AWS IAM, identifying five distinct PE categories that share a unifying trait: all PEs arise from permissions \emph{propagating} across entities. We formalize this as \emph{permission flows} and manually extract 219 permission flow templates from a systematic review of all AWS-supported operations (14,000+), covering all five categories. Building on this abstraction, we develop TAC-WB, a whitebox detector that achieves comprehensive PE coverage. To address the second limitation (full configuration access required), we develop TAC-GB, the first greybox PE detector, which operates on partial configurations where customers freely choose which entities to disclose and selectively accept or decline queries about permission assignments among them. A dynamic query mechanism adapts the detection trajectory based on each customer response, with reinforcement learning over graph neural networks selecting the most informative queries to minimize customer interactions.
Furthermore, to support training and evaluation, we construct TAC-Bench, the first comprehensive IAM PE benchmark comprising 2,500 tasks that match the scale, complexity, and diversity of real-world misconfigurations. 

Experiments across synthetic, public, and real-world benchmarks show that TAC-WB detects all PEs missed by state-of-the-art tools, including real-world enterprise escalations where every existing detector fails, while maintaining practical runtime. TAC-GB outperforms all greybox variants in both effectiveness and query efficiency, often rivaling whitebox baselines that require full configuration access, even under modest query budgets. Looking ahead, TAC can be extended to other cloud platforms such as Google Cloud and Azure, as well as beyond IAM, laying the foundation for cross-platform PE detection and broader greybox security analysis.
\end{abstract}

%% file: intro.tex
\section{Introduction}
\label{sec:intro}

Identity and Access Management (IAM)~\cite{awsiam} is a core access control service in cloud platforms such as Amazon Web Services (AWS). Customers define an \emph{IAM configuration} of entities---users, roles, and services (e.g., EC2 instances)---together with the permissions assigned to them, and IAM evaluates every service request against this configuration. Misconfigurations in IAM can lead to data breaches, denial of service, and resource hijacking~\cite{capitalone,impact1,impact2,impact3,isc2:report21}, with significant economic losses~\cite{dollars}. Among the most dangerous consequences is IAM Privilege Escalation (PE)~\cite{rhino:privesc,bishopstudy,sonraistudy}, where an attacker exploits configuration flaws to acquire permissions beyond those originally granted, gaining unauthorized access to sensitive operations or restricted resources.

To detect IAM PEs, both cloud providers~\cite{awsiamaccess,backes2018semantic,bouchet2020block,awsvp,awsiampolicysimu} and third-party security vendors~\cite{thirdpartywiz,thirdpartyorca,thirdpartysymmetry} have developed detection tools. Among these, third-party services---which typically address more specialized customer requirements than native cloud offerings~\cite{nativevsthirdpartysysdig,nativevsthirdparty}---have produced the most advanced PE detectors~\cite{shevrin2023detecting,yang2023ase,pacu,pmapper,cloudsplaining,awspx}. However, these detectors face two key limitations. 

\emph{\uline{Limitation~1: limited PE coverage.}} Each detector covers only a small subset of PE types, leaving many vulnerabilities undetected. For example, Pacu~\cite{pacu} encodes only 21 known PE patterns~\cite{rhino:privesc}, PMapper~\cite{pmapper} models only user/role authentication chains, and none achieves full coverage across all five PE categories identified in our study. As a result, organizations relying on any single detector, or even a combination of existing tools, remain exposed to entire classes of PEs.

\emph{\uline{Limitation~2: full IAM configuration access required.}} All existing detectors require full access to IAM configurations, operating exclusively as whitebox tools. This poses a significant barrier for third-party security services, whose customers in security-sensitive domains (e.g., healthcare, finance, government~\cite{oqaily2019segguard,kolhar2017cloud,ryoo2013cloud}) are often unwilling to disclose complete configurations for fear of exposing sensitive organizational details. Industry surveys report that over 90\% of organizations express concerns about sharing cloud configuration data with external parties~\cite{isc2:report21}, and cloud security best practices explicitly advocate minimizing configuration exposure~\cite{awsbestpractice}. In practice, customers must manually anonymize sensitive information before sharing---a process that is both time-consuming and error-prone, fundamentally limiting the applicability of existing detection approaches.

To overcome Limitation~1 (limited PE coverage), we first conduct a systematic study of permission acquisition in cloud IAM, identifying five distinct PE categories (Section~\ref{sec:study}). A key finding is that all five share a common trait: permissions \emph{propagate} across entities---through policy attachment, role assumption, group membership, code execution, or credential manipulation. We formalize this as \emph{permission flows} and manually extract 219 \emph{permission flow templates} from a systematic review of the complete AWS operation set~\cite{awsactiontable} (14,000+ operations), covering all five categories. We then develop TAC-WB, a whitebox detector that encodes IAM configurations as a \emph{Permission Flow Graph (PFG)} and applies domain-specific fixed-point analysis to systematically trace permission propagation, achieving comprehensive PE coverage.

To address Limitation~2 (full IAM configuration access required), we develop TAC-GB, the first greybox PE detector. TAC-GB builds on the insight that detecting a PE requires only the permissions along the relevant propagation paths, not the entire configuration. Customers may disclose any subset of entities and accept or decline each query about permission assignments among them---retaining full control over what they reveal. Given such a partial configuration, TAC-GB extends the PFG to a \emph{partial PFG} that encodes what is currently known and interactively queries customers through a dynamic mechanism that adapts the detection trajectory based on each response. To further reduce customer effort, TAC-GB employs reinforcement learning (RL) over graph neural networks (GNNs) to learn which queries are most informative, minimizing the number of interactions needed to detect PEs.

Training TAC-GB and evaluating both detectors require a large, diverse corpus of IAM PE tasks, yet the only public dataset (\texttt{IAM Vulnerable}~\cite{iamvulnerable}) contains just 31 simple tasks. We therefore construct TAC-Bench, the \emph{first comprehensive IAM PE benchmark}, comprising 2,500 tasks designed to mimic the scale, diversity, and complexity of real-world IAM misconfigurations. Tasks are synthesized via a structured three-stage LLM pipeline---realistic context generation, PE generation grounded in documented real-world attack techniques~\cite{awsactiontable,rhino:privesc,bishopstudy,sonraistudy,capitalonereport}, and chain-of-thought verification---with every configuration manually validated. TAC-Bench configurations average 137 entities and 1,282 permissions, closely matching the scale of real enterprise deployments.  

Experiments across synthetic, public, and real-world benchmarks demonstrate that TAC-WB detects all PEs that state-of-the-art tools miss---including real-world enterprise escalations where every existing detector fails---while maintaining practical runtime. TAC-GB, under modest query budgets, outperforms all greybox variants in both effectiveness and query efficiency, often rivaling whitebox baselines that require full configuration access.

The contributions of this paper are: 
\begin{itemize} [nosep, left=0pt]
\item \textbf{Study.} A systematic study of permission acquisition in cloud IAM that identifies five distinct PE categories and reveals the coverage gaps of all existing detectors.
\item \textbf{Whitebox PE Detector.} TAC-WB, which leverages 219 systematically extracted permission flow templates and fixed-point analysis over a Permission Flow Graph to achieve comprehensive PE coverage.
\item \textbf{Greybox PE Detector.} TAC-GB, the first greybox PE detector, which extends the PFG to a partial PFG and detects PEs from partial configurations through interactive customer queries guided by RL-trained GNNs. 
\item \textbf{Dataset.} TAC-Bench, the first comprehensive IAM PE benchmark with 2,500 tasks that match the scale, complexity, and diversity of real-world enterprise configurations.
\end{itemize}

%% file: background.tex
\section{Background: IAM Basics}

\label{sec:iam_basics}
\noindent\textbf{IAM Configurations.}
\label{sec:back_iam_config} 
An IAM configuration can be viewed as a set of relations among entities and permissions. Entities represent subjects (e.g., users, user groups, services) or roles. Subjects actively perform actions, while roles encapsulate job functions and responsibilities within an organization. Permissions define privileges to perform operations and can be assigned to entities. 

IAM configurations include two types of relations: entity–permission and entity–entity. Entity–permission relations specify which permissions are directly assigned to which entities. Entity–entity relations capture how entities are associated with one another, as defined in official cloud service documentation. Taken together, these relations can be modeled as a multi-relational graph: nodes represent entities and permissions, and edges represent either entity–permission or entity–entity associations.

\begin{figure}[t]
	\begin{subfigure}[b]{0.47\columnwidth}
		\centering
		\includegraphics[width=\textwidth]{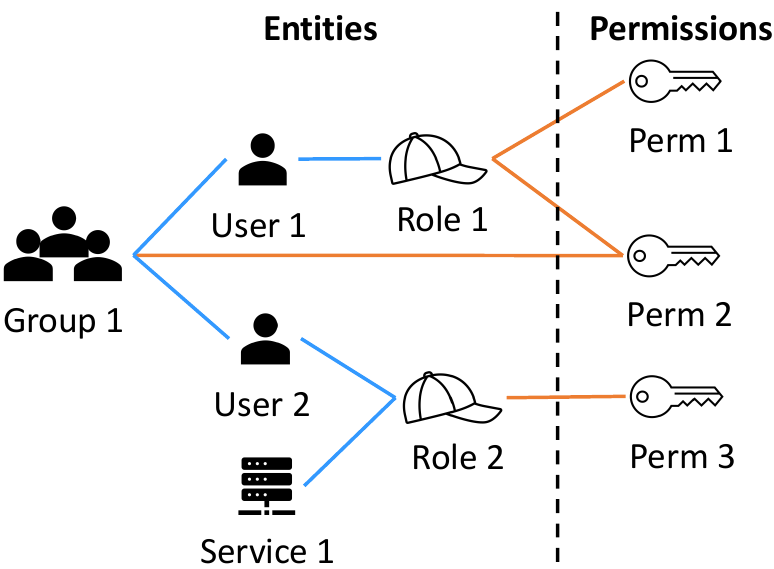}
		\caption{IAM configuration example.}
		\label{fig:iam:rbac}
	\end{subfigure}
	\hfill
	\begin{subfigure}[b]{0.47\columnwidth}
		\centering
        \includegraphics[width=\textwidth]{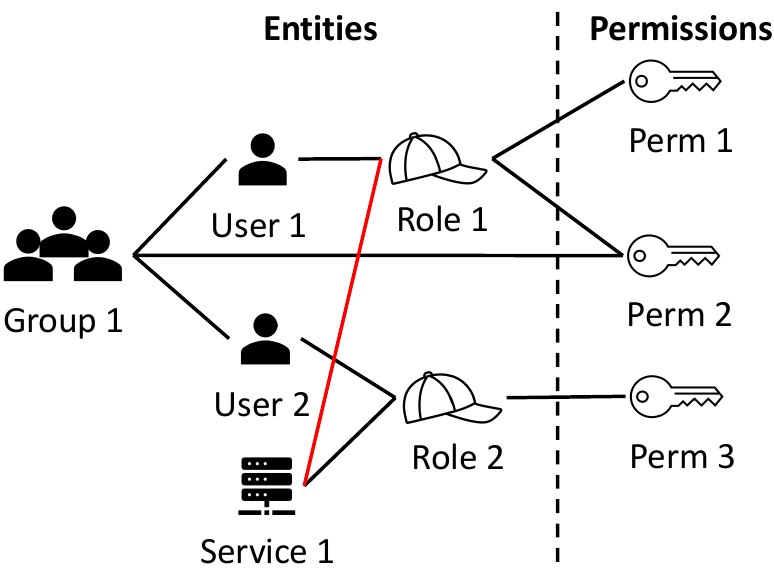}
		\caption{Modified configuration in PE.}
		\label{fig:priv-escalation}
	\end{subfigure}
	\caption{Illustration of IAM configuration.}
\end{figure}

Figure~\ref{fig:iam:rbac} illustrates such a graph. The example includes six entities of four types: one user group (\texttt{Group 1}), two users (\texttt{User 1}, \texttt{User 2}), one service (\texttt{Service 1}), and two roles (\texttt{Role 1}, \texttt{Role 2}); along with three permissions: \texttt{Perm 1}, \texttt{Perm 2}, and \texttt{Perm 3}. Orange edges denote entity-permission relations (direct permission assignments): \texttt{Perm 1} and \texttt{Perm 2} are directly assigned to \texttt{Role 1}, \texttt{Perm 2} to \texttt{Group 1}, and \texttt{Perm 3} to \texttt{Role 2}. Blue edges denote entity–entity relations, such as user–group (users belonging to a group), user–role (a user assuming a role), and service–role (a service assuming a role). Through these relations, permissions can also be inherited indirectly. For example, \texttt{User 1} and \texttt{User 2}, as members of \texttt{Group 1}, inherit \texttt{Perm 2}, while \texttt{User 2} and \texttt{Service 1}, by assuming \texttt{Role 2}, inherit \texttt{Perm 3}.

\noindent\textbf{PEs due to Misconfigurations.}
\label{subsec:pe_iam}
PE in IAM~\cite{rhino:privesc,bishopstudy,sonraistudy} occurs when an attacker exploits IAM misconfigurations to gain unauthorized permissions for sensitive operations or restricted resources. Attackers can exploit these misconfigurations by altering static IAM policy documents or dynamically acquiring new permissions (e.g., by assuming a role). We define the \emph{untrusted entity} as the attacker-controlled entity and the \emph{target permission} as the permission the attacker seeks. A PE is successful when the attacker leverages misconfigurations to assign the target permission to the untrusted entity.

\begin{table*}[t!]
    \centering
    \scalebox{0.9}{
    \begin{tabular}{|l|c|c|c|c|c|}
    \hline
    \textbf{Category} & \textbf{TAC} & \textbf{Pacu} \cite{pacu} & \textbf{Cloudsplaining} \cite{cloudsplaining} & \textbf{PMapper} \cite{pmapper} & \textbf{Shevrin\&Margalit} \cite{shevrin2023detecting} \\
    \hline
    Permission Request & \cmark & \pmark & \pmark & \pmark & \pmark \\
    \hline
    Permission Delegation & \cmark & \pmark & \pmark & \xmark & \xmark \\
    \hline
    Permission Inheritance & \cmark & \pmark & \pmark & \pmark & \pmark \\
    \hline
    Code Execution & \cmark & \pmark & \pmark & \xmark & \pmark \\
    \hline
    Credential Takeover & \cmark & \pmark & \pmark & \xmark & \pmark \\
    \hline
    \end{tabular}
    }
    \caption{Overview of five PE categories covered by SOTA PE detectors. \cmark~means ``fully covered'', \pmark~means ``partially covered'', and \xmark~means ``not covered''.}
    \label{tab:permflow-categories}
    \end{table*}

	\begin{table*}[t!]
\centering
\scalebox{0.9}{
\begin{tabular}{|c|c|c|c|c|c|c|}
\hline
\shortstack[c]{\textbf{Permission Acquisition}\\\textbf{Category}} &
\shortstack[c]{\textbf{Permission}\\\textbf{Request}} &
\shortstack[c]{\textbf{Permission}\\\textbf{Delegation}} &
\shortstack[c]{\textbf{Permission}\\\textbf{Inheritance}} &
\shortstack[c]{\textbf{Code}\\\textbf{Execution}} &
\shortstack[c]{\textbf{Credential}\\\textbf{Takeover}} &
\shortstack[c]{~\\\textbf{Total}\\~} \\\hline
\textbf{\# Template} & 15 & 149 & 6 & 42 & 7 & 219 \\\hline
\end{tabular}
}
\caption{Permission flow templates for five PE categories.}
\label{tab:templatescount}
\end{table*}

Figure~\ref{fig:iam:rbac} illustrates our running example of an IAM misconfiguration \emph{derived from the 2019 Capital One data breach}~\cite{capitalone,khan2022systematic}---one of the most impactful cloud security incidents, which exposed over 100 million customer records. In the real attack, an adversary exploited an overly permissive IAM role on an EC2 instance to assume a second role with S3 access, exfiltrating sensitive data. Our example faithfully models this scenario: \texttt{Service~1} is the attacker-controlled EC2 instance (the untrusted entity), \texttt{Role~2} is its execution role, and \texttt{Role~1} holds the target S3 permission.
The root cause is \texttt{Perm~3} (assigned to \texttt{Role~2}), which grants the ability to assume \texttt{Role~1}. The PE unfolds as follows: \texttt{Service~1} inherits \texttt{Perm~3} by assuming \texttt{Role~2}, then uses it to assume \texttt{Role~1}, thereby obtaining the target permission \texttt{Perm~1} (\texttt{s3:GetObject}).

%% file: motivation.tex
\section{Study of IAM Privilege Escalation}
\label{sec:study}
To overcome Limitation~1 (limited PE coverage), we first conduct a systematic study of how permissions can be acquired in cloud IAM. We draw from the complete AWS operations table~\cite{awsactiontable} (14,000+ operations), official AWS documentation~\cite{awsbestpractice,awsiam}, and well-established prior studies on IAM PE~\cite{rhino:privesc,bishopstudy,sonraistudy,capitalonereport}. Our analysis identified five distinct \emph{PE categories}, each exploiting a fundamentally different mechanism by which an entity can acquire additional permissions: 
\begin{itemize} [nosep, left=0pt]
\item \emph{\textbf{Permission Request: }} Submit a request to cloud access control to grant additional permissions to an entity. For example, an entity may request that an IAM policy be attached to itself via \texttt{iam:AttachUserPolicy} to get more permissions. 
\item \emph{\textbf{Permission Delegation: }} Use an entity as a delegate to request that the cloud access control grant permissions to another entity. For instance, updating a CloudFormation stack with the operation \texttt{cloudformation:UpdateStack} can modify a Lambda function's execution role, thereby granting that entity new permissions. 
\item \emph{\textbf{Permission Inheritance: }} Acquire permissions by establishing a permission inheritance relationship with another entity that already holds them. For example, adding a user entity to an IAM group with \texttt{iam:AddUserToGroup} confers the group's permissions on that entity. 
\item \emph{\textbf{Code Execution: }} Execute code under the security context of an entity. For example, updating the code of a Lambda function through \texttt{lambda:UpdateFunctionCode} can run code with the permissions of the Lambda function. 
\item \emph{\textbf{Credential Takeover: }} Create or reset credentials for an entity in order to act with its permissions. For example, creating an access key (\texttt{iam:CreateAccessKey}) or resetting a login profile (\texttt{iam:UpdateLoginProfile}) enables acting as that entity with its full permissions.
\end{itemize}

We then assessed how well state-of-the-art detectors---Pacu~\cite{pacu}, Cloudsplaining~\cite{cloudsplaining}, PMapper~\cite{pmapper}, and the approach by Shevrin and Margalit~\cite{shevrin2023detecting}---cover these five categories. As Table~\ref{tab:permflow-categories} shows, the coverage gaps are significant: \emph{no detector fully covers even a single category}. Worse, PMapper misses three categories entirely (Permission Delegation, Code Execution, and Credential Takeover), and Shevrin and Margalit completely miss Permission Delegation. Even Pacu and Cloudsplaining, which partially touch all five categories, cover only a small fraction of the operations within each. These findings confirm that Limitation~1 is not merely a matter of missing a few edge cases---existing tools systematically fail to address large portions of the PE attack surface.

%% file: whitebox.tex
\section{TAC-WB}
\label{sec:tac-wb}

\begin{figure*}[t]
\centering
	\begin{subfigure}{0.24\textwidth}
		\includegraphics[width=\linewidth]{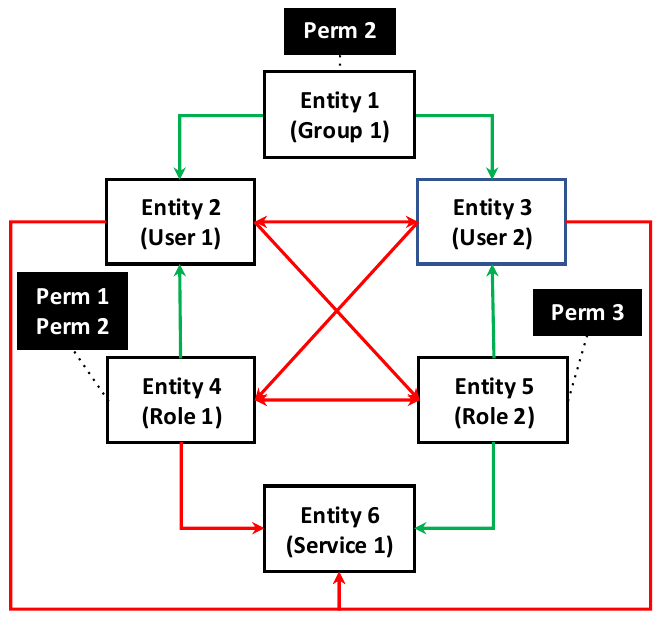}
		\caption{PFG.}
		\label{fig:pfg:direct}
	\end{subfigure}
	\begin{subfigure}{0.24\textwidth}
		\includegraphics[width=\linewidth]{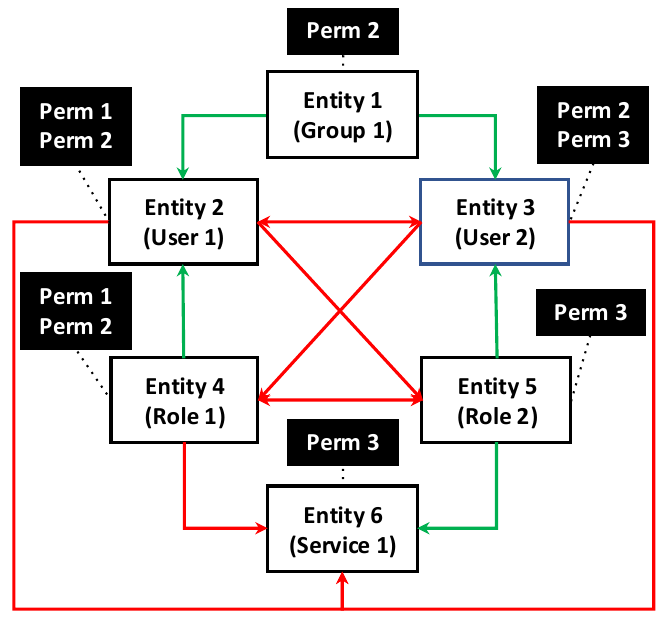}
		\caption{Semantic Representation.}
		\label{fig:pfg:prop}
	\end{subfigure}
	\begin{subfigure}{0.24\textwidth}
		\includegraphics[width=\linewidth]{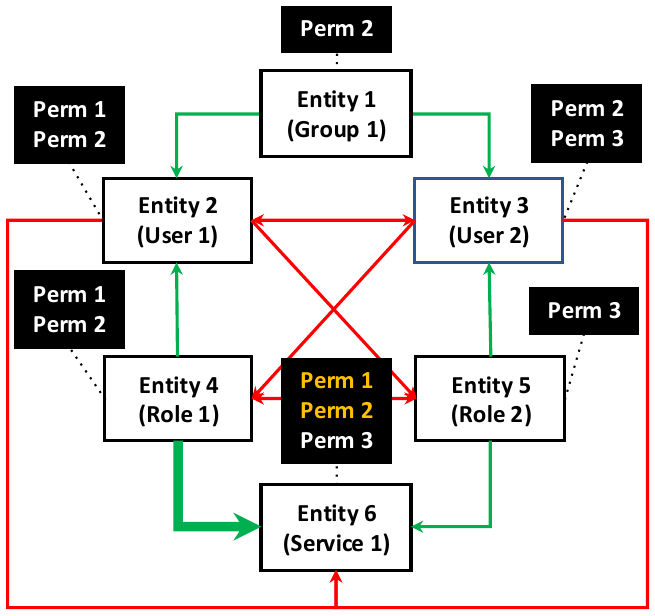}
		\caption{Updated PFG.}
		\label{fig:pfg:g1}
	\end{subfigure}
	\begin{subfigure}{0.24\textwidth}
		\includegraphics[width=\linewidth]{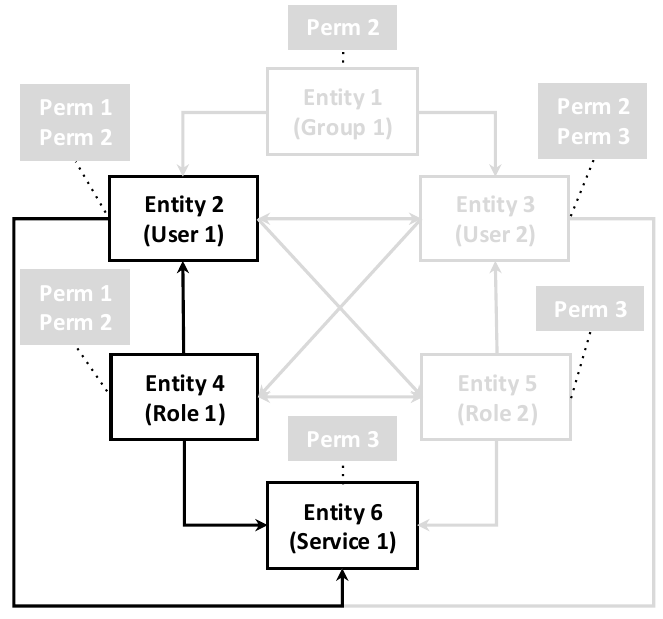}
		\caption{Partial PFG.}
		\label{fig:pfg:visible}
	\end{subfigure}
	\caption{PFGs of the example in Figure~\ref{fig:iam:rbac}; Green indicates enabled permission flows; red indicates disabled permission flows.}
	\label{fig:pfg}
\end{figure*}

We propose TAC-WB, a whitebox detector designed to achieve comprehensive coverage across all five PE categories identified in our study. TAC-WB exploits the key observation that all five PE categories share a common trait: permissions \emph{propagate} across entities---whether through policy attachment, role assumption, group membership, code execution, or credential manipulation. This observation guides our design of TAC-WB as follows.

We first define \emph{permission flows} to model how permissions propagate between entities, and systematically extract 219 \emph{permission flow templates} covering all five PE categories. From these templates, we derive the \emph{permission space}---the subset of permissions that can contribute to PEs---filtering out irrelevant permissions (Section~\ref{subsec:PFModeling}). We then use these to model IAM configurations by encoding the flows into a \emph{Permission Flow Graph} and computing the \emph{semantic representation} of the configuration through domain-specific fixed-point analysis (Section~\ref{subsec:IAMModeling}).  Finally, TAC-WB reports a PE only when the semantic representation confirms that the untrusted entity can reach the target permission through a sequence of permission propagations, ensuring precision by construction---every reported PE corresponds to a concrete escalation path (Section~\ref{subsec:wb-detection}).

\subsection{Permission Propagation Modeling} 
\label{subsec:PFModeling}
\subsubsection{Permission Flow}
An entity can gain permissions in two ways: \emph{direct assignment}, by modifying an entity--permission relation (e.g., attaching a policy to a user), or \emph{indirect assignment}, by modifying an entity--entity relation (e.g., adding a user to a group so the user inherits the group's permissions). For modeling uniformity, we represent all direct assignments as indirect ones by introducing a pseudo-entity for each permission set; assigning permissions to an entity then becomes creating a relation to the corresponding pseudo-entity. This allows us to reason exclusively in terms of indirect assignments.

We define a \emph{permission flow} as an indirect permission assignment from entity $e_1$ to entity $e_2$ (including those derived from direct ones). Each flow has a state---enabled or disabled---indicating whether permissions of $e_1$ currently propagate to $e_2$. A flow exists whenever the relationship between two entity types supports indirect assignment. For instance, the user--role relation means all permissions of a role propagate to any user assuming it. Figure~\ref{fig:pfg:direct} illustrates the resulting flows for the configuration in Figure~\ref{fig:iam:rbac}: the flow from \texttt{Role~1} to \texttt{User~1} is enabled, while the flow from \texttt{Role~2} to \texttt{User~1} is disabled.

\myparagraph{219 Manually Extracted Permission Flow Templates}
\label{subsec:PFT}
To systematically capture which entity relations support permission flows, we manually reviewed the official documentation for every AWS-supported operation~\cite{awsactiontable}, examining whether each operation can enable a permission flow between specific entity types. We clustered operations sharing the same source--sink entity-type pair into \emph{permission flow templates}---each a triple \emph{(source entity type, sink entity type, set of cloud operations)} specifying that any source-type entity can establish a flow to any sink-type entity via the listed operations. For example, the template \texttt{(Group, User, \{\texttt{iam:AddUserToGroup}\})} specifies that a group-to-user flow can be created by invoking \texttt{iam:AddUserToGroup}.

This effort produced \uline{219 permission flow templates} spanning all five PE categories (Table~\ref{tab:templatescount}). These templates directly address Limitation~1: whereas existing detectors encode at most a few dozen ad-hoc patterns covering only some categories (Section~\ref{sec:study}), our 219 templates are derived from a systematic review of the full AWS operation set, offering far broader coverage. All subsequent steps of TAC-WB build on these templates.

\subsubsection{Permission Space}
Permission flows capture how permissions propagate, but propagation can also be triggered: certain permissions, when held by an entity, can enable currently disabled flows and open new propagation paths. For each permission in the IAM configuration, we extract its associated cloud operation and check whether it appears in any template; if so, the template indicates which flows that permission can enable. We define the \emph{permission space} as the union of the target permission and all such flow-enabling permissions. Permissions outside this space cannot contribute to PEs and are excluded, focusing the analysis on the relevant subset.

For example, in Figure~\ref{fig:iam:rbac}, \texttt{Perm~1} is the target permission and belongs to the permission space by definition. \texttt{Perm~2} and \texttt{Perm~3} allow their assignees to assume \texttt{Role~2} and \texttt{Role~1}, respectively; since our templates include role-assumption operations, both are also part of the permission space.

\subsection{IAM Configuration Modeling}
\label{subsec:IAMModeling}
\subsubsection{Permission Flow Graph}
Using the permission flows and permission space, we construct a graph that encodes the full permission propagation structure of an IAM configuration. A \emph{Permission Flow Graph (PFG)} is defined as \( G = (E, F, \mathcal{A}, \mathcal{W}) \), where \( E \) is the set of entities, \( F \subseteq E \times E \) is the set of permission flows, \( \mathcal{A}: E \to 2^P \) maps each entity to its assigned permissions in the permission space \( P \), and \( \mathcal{W}: F \to \{\texttt{true}, \texttt{false}\} \) indicates whether each flow is enabled or disabled.

Figure~\ref{fig:pfg:direct} illustrates a PFG which straightforwardly models the IAM configuration in Figure~\ref{fig:iam:rbac}. Three entities (i.e., \texttt{Group 1}, \texttt{Role 1}, and \texttt{Role 2}) have directly assigned permissions. Based on the semantics of user-role, service-role, and user-group relations, each associated with indirect permission assignments, 12 permission flows are identified. As shown in Figure \ref{fig:iam:rbac}, five of the entity-entity pairs are currently related, thus resulting in five enabled flows, while the remaining seven are disabled.

\subsubsection{Semantic Representation of IAM Configuration}
A PFG captures the \emph{static} structure of an IAM configuration---which flows exist and which are currently enabled. To reason about PE, we need to determine which permissions each entity \emph{effectively} holds after propagation through all enabled flows. We compute this by iteratively applying a permission flow function \( \mathcal{M} \) that propagates permissions along enabled flows. Given a PFG \( G = (E, F, \mathcal{A}, \mathcal{W}) \), \( \mathcal{M} \) produces \( G' = (E, F, \mathcal{A}', \mathcal{W}) \) where:

\begin{equation}
\notag
\mathcal{A}'(e_2) = \bigcup_{\{(e_1, e_2) \in F \mid \mathcal{W}(e_1, e_2) = \texttt{true}\}} \mathcal{A}(e_1) \cup \mathcal{A}(e_2),
\end{equation}
meaning that each entity \( e_2 \) receives all permissions from entities with enabled flows to it. We apply \( \mathcal{M} \) iteratively until a fixed point is reached; the result, \( G' = \mathcal{M}^*(G) \), is the \emph{semantic representation} of the IAM configuration.

Figure~\ref{fig:pfg:prop} illustrates this for our running example. Through the five enabled flows: (1)~\texttt{User~1} and \texttt{User~2} inherit \texttt{Perm~2} from \texttt{Group~1}; (2)~\texttt{User~2} and \texttt{Service~1} inherit \texttt{Perm~3} from \texttt{Role~2}; and (3)~\texttt{User~1} inherits \texttt{Perm~1} and \texttt{Perm~2} from \texttt{Role~1}. No further propagation is possible, confirming the fixed point.

\subsection{Whitebox PE Detection}
\label{subsec:wb-detection}
An IAM configuration has a PE if and only if the untrusted entity can obtain the target permission through its assigned permissions. Formally, let \( G_0 = (E_0, F_0, \mathcal{A}_0, \mathcal{W}_0) \) represent the semantic representation of the initial IAM configuration, \( u \) the untrusted entity, and \( l \in P \setminus \mathcal{A}_0(u) \) the target permission. We define that \( G_0 \) has a PE iff there exists a sequence of permissions \( p_1, \ldots, p_n \) with \( p_i \in \mathcal{A}_{i-1}(u) \), and an updated configuration \( G_n = (E_n, F_n, \mathcal{A}_n, \mathcal{W}_n) \) such that

\begin{equation}
\notag
G_0 \overset{p_1}{\hookrightarrow} G_1 \ldots \overset{p_n}{\hookrightarrow} G_n \land l \in \mathcal{A}_n(u),
\end{equation}
where \( G_{i-1} \overset{p_i}{\hookrightarrow} G_i \) indicates that the configuration \( G_{i-1} \) is updated to \( G_i \) by the untrusted entity \( u \) using the permissions \( p_i \in \mathcal{A}_{i-1}(u) \). \uline{This formulation guarantees precision by construction}: TAC-WB reports a PE only when it finds a concrete sequence of permissions that enables the untrusted entity to reach the target permission, so every reported PE corresponds to a concrete escalation path.

For example, consider the semantic representation of the example IAM configuration in Figure~\ref{fig:pfg:prop}, where \texttt{Service 1} is the untrusted entity and \texttt{Perm 1} the target permission. When \texttt{Service 1} applies \texttt{Perm 3}, the permission flow from \texttt{Role 1} to \texttt{Service 1} is enabled, assigning all permissions of \texttt{Role 1} (i.e., \texttt{Perm 1} and \texttt{Perm 2}) to \texttt{Service 1}. This update is denoted as \( G_0 \overset{\texttt{Perm 3}}{\hookrightarrow} G_1 \), where \( G_0 \) is the semantic representation of the original configuration, and \( G_1 \) (shown in Figure~\ref{fig:pfg:g1}) is the semantic representation of the resulting configuration. Since \texttt{Perm 1} is now assigned to \texttt{Service 1} (\( \texttt{Perm 1} \in \mathcal{A}_1(\texttt{Service 1}) \)), the original configuration is determined to have a PE. 

%% file: greybox.tex
\section{TAC-GB}
\label{subsec:query-abstract-ep}
To address Limitation~2 (full IAM configuration access required), we propose TAC-GB, the first greybox PE detector. The key insight is that detecting a PE does \emph{not} require the entire IAM configuration---it suffices to know the permissions along the propagation paths leading to the target permission. TAC-GB exploits this by allowing customers to freely choose which entities to disclose (with all disclosed entities automatically anonymized) and to accept or decline each query about permission assignments among these disclosed entities, leaving all other configuration details completely hidden. 

TAC-GB realizes this insight in three stages while maintaining \emph{precision by construction}---every reported PE corresponds to a verified escalation path, even under partial visibility. We first extend the PFG to a \emph{partial PFG} that models IAM configurations under limited visibility, where entity permissions and flow states may be unknown (Section~\ref{subsec:partial-iam-modeling}). We then design an interactive query mechanism that probes the customer one question at a time and adapts the detection trajectory based on each response---accepted queries update the semantic representation, while declined queries redirect the search toward alternatives (Section~\ref{subsec:greybox-pe-detection}). To minimize customer effort, we employ reinforcement learning with GNN-based query models that learn to select the most informative queries, reducing the number of interactions needed to detect PEs (Section~\ref{sec:tac:framework}).

\subsection{Partial IAM Configuration Modeling}
\label{subsec:partial-iam-modeling}

\subsubsection{Partial PFG}
\label{subsubsec:abstract_pfg}
We refer to the subset of entities disclosed by the customer as the \emph{visible entities} \( \hat{E} \subseteq E \). From \( \hat{E} \) we derive the \emph{visible permission flows} \( \hat{F} \subseteq F \), flows whose source and target are both in \( \hat{E} \); and the \emph{visible permission space} \( \hat{P} \subseteq P \), comprising the target permission and all permissions that can enable a visible flow. For our running example shown in Figure~\ref{fig:pfg:visible}, the customer discloses \texttt{User~1}, \texttt{Role~1}, and \texttt{Service~1}, and the visible flows are those connecting them.

Using these components, we define a \emph{partial PFG} \( \hat{G} = (\hat{E}, \hat{F}, \hat{\mathcal{A}}, \hat{\mathcal{W}}) \). It mirrors the full PFG (Section~\ref{sec:tac-wb}) but replaces ground-truth values with \emph{visibility states} that evolve dynamically. Two state functions track what is currently known. The entity-permission state function \( \hat{\mathcal{A}}: \hat{E} \times \hat{P} \to \{?,+\} \) assigns \( ? \) when it is unknown whether entity \( e \) holds permission \( p \), and \( + \) when the customer has confirmed the assignment. Likewise, the flow state function \( \hat{\mathcal{W}}: \hat{F} \to \{*, \oplus\} \) assigns \( * \) when it is unknown whether flow \( f \) can be enabled, and \( \oplus \) when the customer has confirmed it can be enabled. 

Initially, no queries have been answered, so all states are unknown: $\forall e\in \hat{E},\, p\in \hat{P}:\; \hat{\mathcal{A}}_{\text{init}}(e, p)=?$ and $\forall f \in \hat{F}:\; \hat{\mathcal{W}}_{\text{init}}(f) = *$. Figure~\ref{fig:aiam:init} shows the resulting initial partial PFG for our running example. As detection proceeds, each accepted query transitions an unknown state (\( ? \) or \( * \)) to a confirmed state (\( + \) or \( \oplus \)), progressively revealing the permission assignments needed to detect PEs.

\begin{figure}[t]
	\centering
	\begin{subfigure}[b]{0.4\columnwidth}
	
	\centering
	\includegraphics[width=0.8\textwidth]{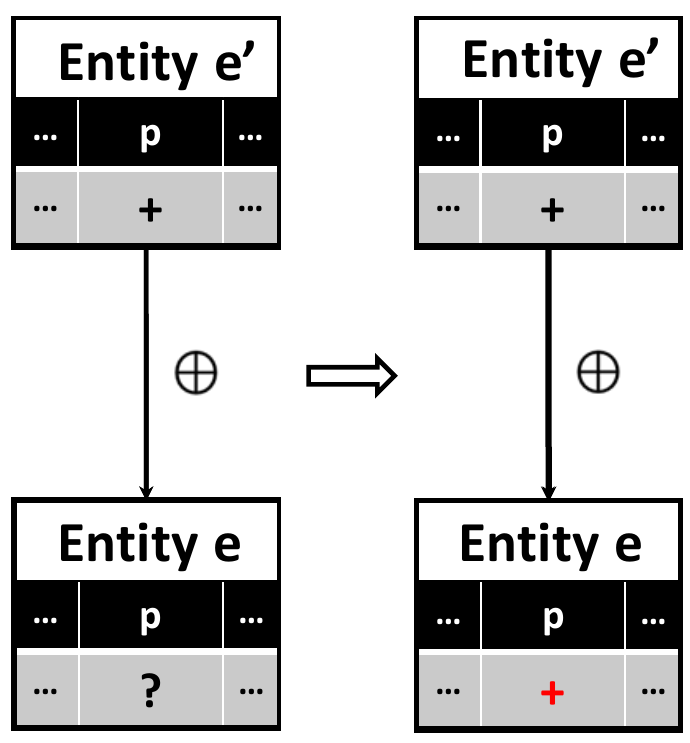}
	\caption{\textbf{C1} update.}

	\label{fig:c1}
\end{subfigure}
		\begin{subfigure}[b]{0.4\columnwidth}
			\centering
		
		\includegraphics[width=0.8\textwidth]{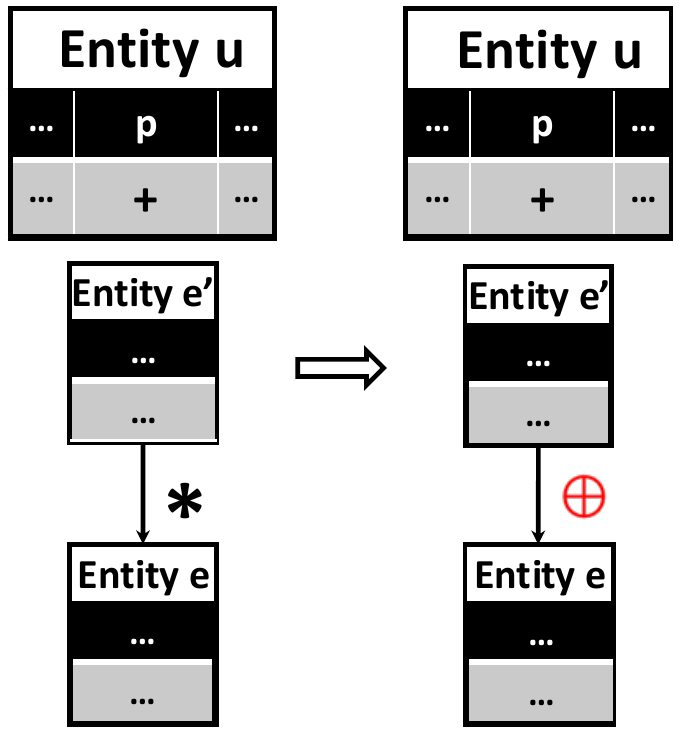}
	\caption{\textbf{C2} update.}
		\label{fig:c2}
	\end{subfigure}
\caption{Visibility state updates.}
\end{figure}
\subsubsection{Semantic Representation of Partial IAM Configuration} 
As the customer confirms permission assignments via queries, these newly revealed permissions can propagate through the visible entities, just as they do in the whitebox setting. To capture the full extent of the untrusted entity's capabilities under partial visibility, we define a permission flow function \( \hat{\mathcal{M}} \) that performs one iteration of permission propagation on a partial PFG \( \hat{G} \), producing an updated partial PFG \( \hat{G}' = \hat{\mathcal{M}}(\hat{G}) \):

\begin{equation}
	\small
	\notag
\hat{\mathcal{A}}'(e,p)=
	\begin{cases}
		+,& \textbf{C1:}~\exists e'\in N_{\oplus}^{(e)}. \hat{\mathcal{A}}(e',p)=+, \\
		\hat{\mathcal{A}}(e, p),& \text{otherwise.}\\
	\end{cases}
\end{equation}

\begin{equation}
	\small
	\notag
	\hat{\mathcal{W}}'(f)=
	\begin{cases} 
		\oplus,& \textbf{C2:}~\exists p\in \hat{P}.\llbracket p \rrbracket=f
		\land \hat{\mathcal{A}}(u,p)=+, \\
		\hat{\mathcal{W}}(f),& \text{otherwise.}\\
	\end{cases}
\end{equation}
Here, $N_{\oplus}^{(e)}=\{e' \mid (e',e)\in \hat{F} \land \hat{\mathcal{W}}(e',e) = \oplus\}$ is the set of entities with a confirmed enabled flow to $e$, and $\llbracket p \rrbracket$ is the flow enabled by permission $p$.
These conditions mirror the whitebox propagation logic but operate strictly on confirmed visibility states. Condition \textbf{C1} (Figure~\ref{fig:c1}) models permission inheritance: if entity $e'$ has a confirmed permission $p$ ($+$) and there is a confirmed enabled flow ($\oplus$) from $e'$ to $e$, then $e$ also acquires $p$, updating its state to $+$. Condition \textbf{C2} (Figure~\ref{fig:c2}) models flow enablement: if the untrusted entity $u$ has a confirmed permission $p$ ($+$) that enables flow $f$, the flow's state updates to $\oplus$.

By repeatedly applying \( \hat{\mathcal{M}} \) until a fixed point is reached (\( \hat{G}' = \hat{\mathcal{M}}^{*}(\hat{G}) \)), we obtain the \emph{semantic representation} of the partial IAM configuration. This representation captures all permissions and flows that are logically deducible from customer responses so far.

\begin{figure}[t]
	\centering
	\begin{subfigure}[b]{\columnwidth}
		\centering
	\includegraphics[width=\textwidth]{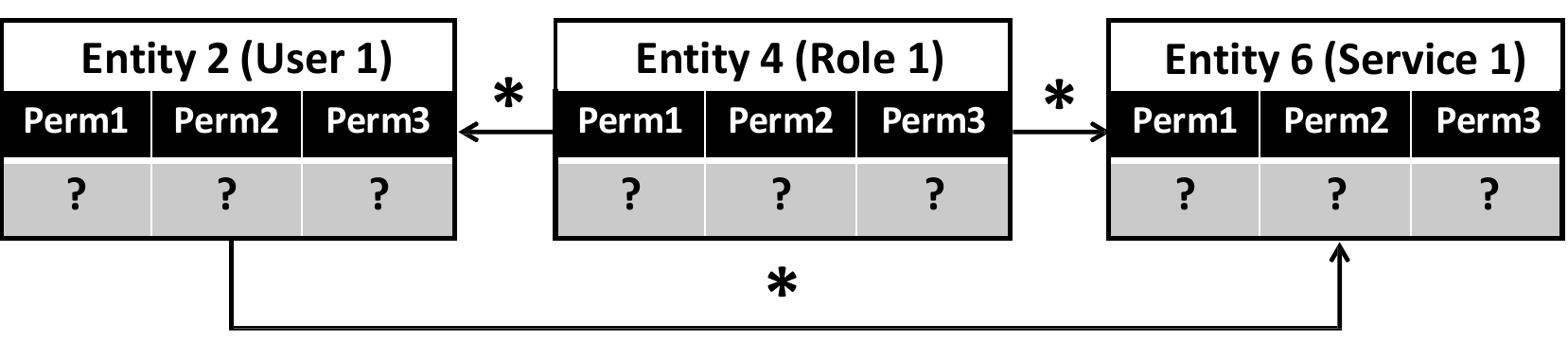}
		\caption{Initial semantic representation of the partial PFG in Figure~\ref{fig:pfg:visible}.}
		\label{fig:aiam:init}
	\end{subfigure}
	\hfill
	\begin{subfigure}[b]{\columnwidth}
		\centering
		\includegraphics[width=\textwidth]{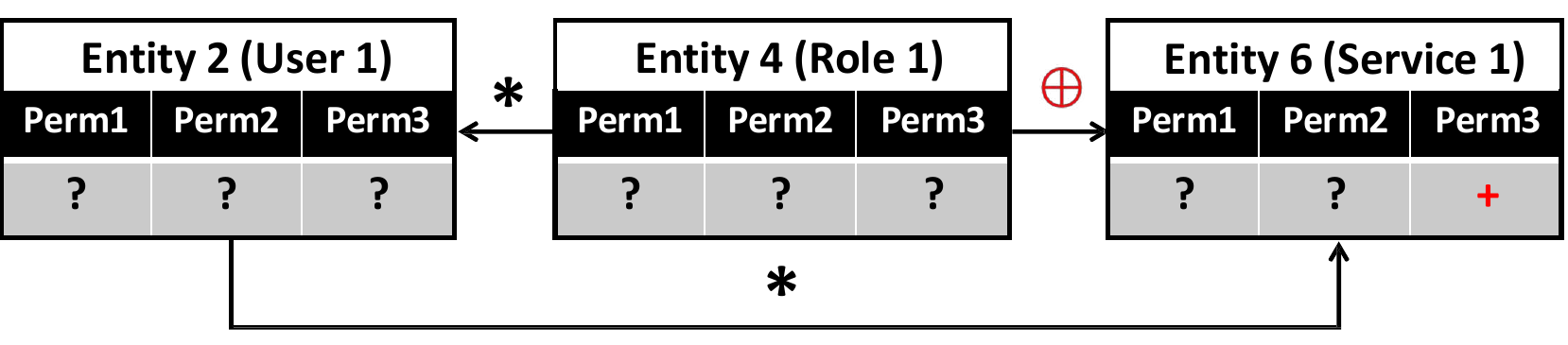}
		\caption{Intermediate representation with response $\mathcal{O}(\texttt{Entity~6}, \texttt{Perm~3})$.}
		\label{fig:aiam:intermediate}
	\end{subfigure}
	\hfill
	\begin{subfigure}[b]{\columnwidth}
		\centering
		\includegraphics[width=\textwidth]{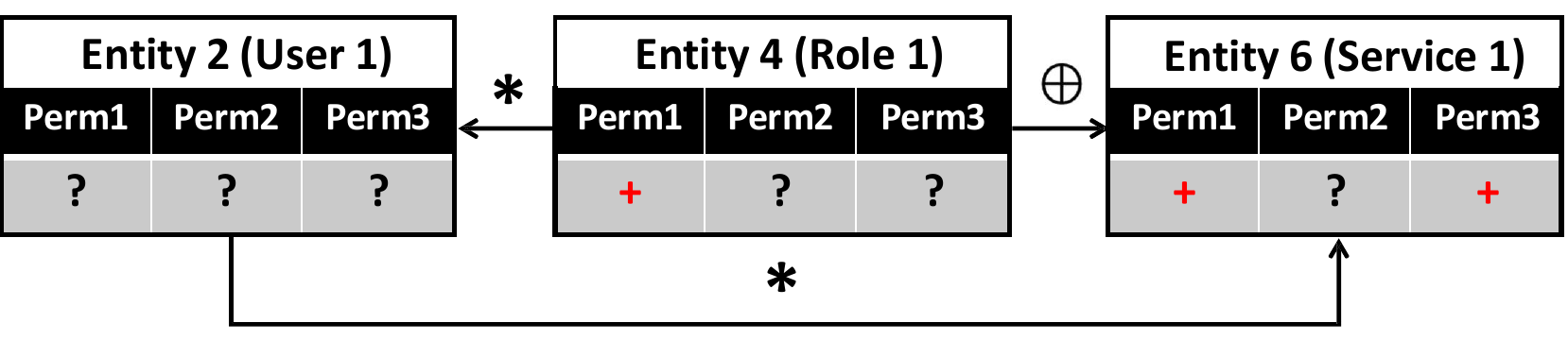}
		\caption{Terminal representation with response  $\mathcal{O}(\texttt{Entity~4}, \texttt{Perm~1})$.}
		\label{fig:aiam:terminal}
	\end{subfigure}
	\caption{An example of greybox PE detection.}
	\label{fig:aiam}
\end{figure}

During the dynamic detection process, this semantic representation continuously evolves. It begins as the \emph{initial} representation \( \hat{G}_{\text{init}} \) (where no queries have been answered). As queries are accepted, it transitions through \emph{intermediate} representations. If the propagation eventually confirms that the untrusted entity \( u \) can reach the target permission \( l \) (i.e., \( \hat{\mathcal{A}}_{\text{term}}(u, l) = + \)), it reaches a \emph{terminal} representation \( \hat{G}_{\text{term}} \), signaling a successfully detected PE. Figure~\ref{fig:aiam} illustrates these three stages for our running example.

\subsection{Greybox PE Detection}
\label{subsec:greybox-pe-detection}
Building on the partial PFG and its semantic representation, we now formalize the greybox PE detection process. The inputs to TAC-GB are the initial semantic representation \( \hat{G}_{\text{init}} \) (constructed from the customer's visible entities), the untrusted entity \( u \), the target permission \( l \), and a \emph{query budget} specifying the maximum number of questions the customer is willing to answer.

The detection proceeds as an interactive, dynamic process. In each step, TAC-GB selects a query \( (e, p) \in \hat{E} \times \hat{P} \), asking the customer: \emph{"Is permission \( p \) assigned to entity \( e \)?"} The customer can choose to accept or decline the query based on whether the assignment is confidential. If they decline, or if they accept but the answer is \texttt{false}, the semantic representation remains unchanged. However, if they accept and answer \texttt{true}, TAC-GB updates the entity-permission state \( \hat{\mathcal{A}}(e, p) \) to \( + \) and immediately applies the fixed-point permission flow function \( \hat{\mathcal{M}}^{*} \) to propagate this newly confirmed permission. This transitions the partial PFG to a new intermediate representation.

Formally, let \( Q = \hat{E} \times \hat{P} \) be the query space. The customer's response function \( \mathcal{O}: Q \to \{\texttt{true}, \texttt{false}, \texttt{unknown}\} \) is defined as:
\begin{equation}
	\notag
	\mathcal{O}(e,p)=
	\begin{cases}
		\texttt{true},& \text{accepted} \land p\in \mathcal{A}(e), \\
			\texttt{false},& \text{accepted} \land p\not\in \mathcal{A}(e), \\
			\texttt{unknown},& \text{declined.}
		\end{cases}	
\end{equation}
When a query \( (e_i, p_i) \) is answered, the semantic representation updates from \( \hat{G}_{i-1} \) to \( \hat{G}_i \):
\begin{equation}
	\notag
	\hat{G}_i=
	\begin{cases}
			\hat{\mathcal{M}}^{*}(\hat{G}_{i-1}[\hat{\mathcal{A}}_{i-1}[e_i,p_i]\mapsto +]),&  \mathcal{O}(e_i,p_i)=\texttt{true}, \\
			\hat{G}_{i-1},& \text{otherwise.}
		\end{cases}	
\end{equation}

TAC-GB successfully detects a PE if, within the query budget, a sequence of queries \( (e_1, p_1), \ldots, (e_n, p_n) \) updates the initial representation \( \hat{G}_0 \) to a terminal representation \( \hat{G}_n \) where the untrusted entity has acquired the target permission:
\begin{equation}
    \notag
\hat{G}_0\overset{(e_1, p_1)}{\hookrightarrow} \hat{G}_1\ldots \overset{(e_n, p_n)}{\hookrightarrow} \hat{G}_n\land \hat{\mathcal{A}}_n(u,l)=+.
\end{equation}
Crucially, because TAC-GB's propagation rules operate strictly on customer-confirmed permissions (\( + \)) and enabled flows (\( \oplus \)), \uline{the greybox detection is precise by construction}. Any PE reported by TAC-GB corresponds to a concrete, verified escalation path, guaranteeing that no false positives are introduced by the partial visibility.

We use the example in Figure~\ref{fig:aiam} to illustrate this process. Starting with the initial representation (Figure~\ref{fig:aiam:init}), TAC-GB queries whether \texttt{Entity~6} has \texttt{Perm~3}. The customer accepts and responds \texttt{true} (\( \mathcal{O}(\texttt{Entity~6}, \texttt{Perm~3}) = \texttt{true} \)). TAC-GB updates the state of (\texttt{Entity~6}, \texttt{Perm~3}) to \( + \). Because \texttt{Perm~3} enables the flow from \texttt{Entity~4} to \texttt{Entity~6}, the fixed-point function \( \hat{\mathcal{M}}^{*} \) automatically updates this flow's state to \( \oplus \), yielding the intermediate representation in Figure~\ref{fig:aiam:intermediate}.
Next, TAC-GB queries whether \texttt{Entity~4} has the target permission \texttt{Perm~1}. The customer again responds \texttt{true}. TAC-GB updates (\texttt{Entity~4}, \texttt{Perm~1}) to \( + \). Through condition \textbf{C1} of \( \hat{\mathcal{M}}^{*} \), this permission flows across the newly enabled \( \oplus \) edge, updating (\texttt{Entity~6}, \texttt{Perm~1}) to \( + \). Since \texttt{Entity~6} (the untrusted entity) has now reached \texttt{Perm~1} (the target permission), the configuration reaches the terminal state (Figure~\ref{fig:aiam:terminal}), and the PE is detected.

\begin{figure}[t]
	\centering
	\includegraphics[width=0.8\columnwidth]{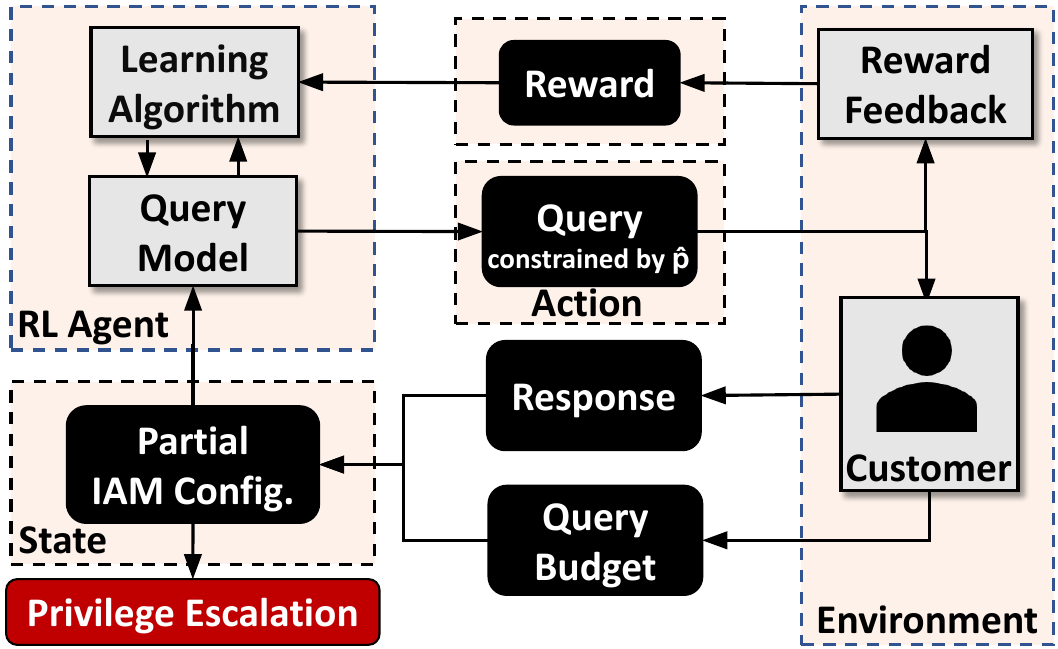}
	\caption{The general framework of TAC-GB.}
	\label{fig:overview}
\end{figure}

\subsection{Optimizing Greybox PE Detection}
\label{sec:tac:framework}
While the interactive process guarantees precise detection, its real-world practicality hinges on minimizing the customer's interaction effort. Because the query space \( \hat{E} \times \hat{P} \) can be massive, naively querying the customer would quickly exhaust their limited query budget and overwhelm security teams with manual configuration inspections. Therefore, TAC-GB must intelligently select the queries, aiming to detect PEs with as few human interactions as possible.

This query selection is inherently a \emph{sequential decision-making} problem: the optimal next query depends entirely on the current semantic representation, which evolves based on previous responses. Returning to our running example (Figure~\ref{fig:aiam}), if the customer confirms that \texttt{Entity~6} has \texttt{Perm~3}, the flow to \texttt{Entity~6} is enabled, making it highly strategic to next ask if \texttt{Entity~4} has the target \texttt{Perm~1}. However, if the customer declines or denies the first query, the flow remains disabled; asking about \texttt{Entity~4} would be a wasted query, and TAC-GB must pivot to explore alternative escalation paths.

To solve this sequential decision-making problem and minimize customer interactions, TAC-GB formulates query selection as a Reinforcement Learning (RL) task. 
	\begin{figure*}[t]
	\centering
	\includegraphics[width=0.75\textwidth]{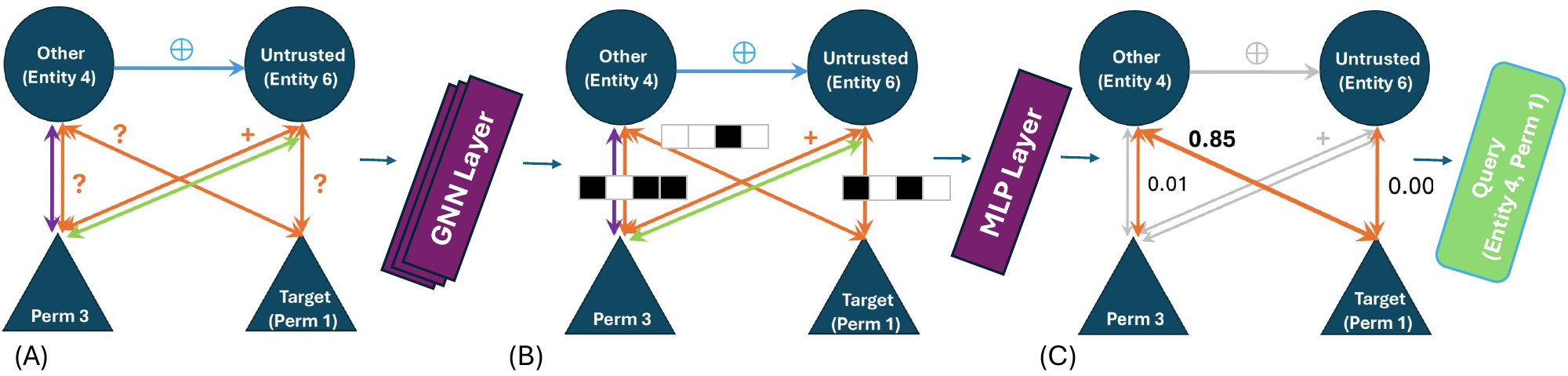}
	\caption{The design and workflow of the query model. (A) shows input graph representation of the IAM configuration for the GNN, converted from the semantic representation in Figure~\ref{fig:aiam:intermediate}. Entities are shown as circles and permissions as triangles. Permission flows are represented by blue directed edges. All other edges are bidirectional: orange edges connect a permission to entities that may own it; purple edges connect a permission to the source entity of the flow it enables; and green edges connect it to the sink entity of that flow. }
	\label{fig:gnnmodel}
\end{figure*}

\subsubsection{RL Formulation}
\label{sec:rl:formulation}
Figure~\ref{fig:overview} illustrates how TAC-GB formulates this sequential decision-making problem as an RL task. In each episode, the RL agent observes the current semantic representation of the partial IAM configuration (the \emph{state}) and uses a query model (the \emph{policy}) to select a specific entity-permission query (the \emph{action}). The environment simulates the customer's response, updates the semantic representation to the next state, and issues a reward of \(-1\). This negative reward per query drives the agent's core objective: to maximize cumulative reward by finding a PE (reaching the terminal state) in as few steps as possible.

A critical design choice that makes this RL formulation tractable is the use of the \emph{visible permission space} $\hat{P}$ (Section~\ref{subsec:partial-iam-modeling}) to constrain the action space. Instead of allowing the agent to query any of the thousands of possible AWS permissions, the action space is strictly limited to the target permission and the specific permissions that can enable visible flows. This domain-specific constraint drastically reduces the search space, forcing the agent to explore only PE-relevant propagation paths within the partial IAM configuration.
\subsubsection{GNN-based Query Model}
\label{sec:rl:model}
To select the query, the RL agent must deeply understand the current state of the configuration to predict which unknown permission assignment is critical for revealing a PE. Because our semantic representations are inherently graph-structured (as partial PFGs), we parameterize our query model using Graph Neural Networks (GNNs). GNNs naturally generalize across diverse graph topologies---essential for learning query strategies that apply to entirely new, unseen customer configurations---and scale efficiently to process the millions of nodes and edges found in real-world enterprise IAM configurations.

\noindent\textbf{Graph Representation.} 
To serve as input for the GNN, the IAM semantic representation is transformed into a directed graph (Figure~\ref{fig:gnnmodel} (A) shows the graph for the intermediate state in Figure~\ref{fig:aiam:intermediate}). The nodes are classified into entities (untrusted vs.\ other) and permissions (target vs.\ flow-enabling). The edges capture the relationships: unidirectional edges between entities represent permission flows, while bidirectional edges between entities and permissions represent assignments or flow requirements. Crucially, these edges carry their current visibility values (\( \hat{\mathcal{W}} \) and \( \hat{\mathcal{A}} \)). Each permission assignment edge with an unknown visibility state (\( ? \)) represents a potential query the agent can select.

\noindent\textbf{Model Architecture.} 
Figure~\ref{fig:gnnmodel} illustrates the workflow of the query model. Given the graph input \texttt{(A)}, five stacked Graph Attention Network (\texttt{GATv2}) layers~\cite{brody2021attentive} process the topology and visibility features to generate rich embeddings for the permission assignment edges \texttt{(B)}. A Multi-Layer Perceptron (MLP) then uses these embeddings to predict a probability distribution specifically over the \emph{query edges}---the permission assignment edges with an unknown state (\( ? \)), as shown in \texttt{(C)}. By selecting the edge with the highest probability, the model outputs the most promising entity-permission query. For example, in Figure~\ref{fig:gnnmodel}, the unknown edge from \texttt{Entity~4} to \texttt{Perm~1} is selected, resulting in the query \( (\texttt{Entity~4}, \texttt{Perm~1}) \).
\subsubsection{Offline Multi-Task Pretraining}
\label{sec:rl:pretraining}
To achieve high efficiency from the very first query, we \emph{pretrain} the GNN query models offline on a diverse set of  PE tasks using multi-task RL. This allows the pretrained model to act effectively in a zero-shot manner when deployed to a new IAM configuration.

To maximize the effectiveness of this pretraining, we must address \emph{negative transfer}---a common issue in multi-task RL where training on conflicting tasks degrades overall performance~\cite{standley2020tasks,rosenstein2005transfer}. We observe that IAM PE strategies differ fundamentally depending on the type of the untrusted entity (e.g., escalating from a \texttt{User} often involves different permission flows than escalating from a \texttt{Service}). Therefore, we partition the PE task space by untrusted entity type (27 types in total), grouping tasks that share similar optimal query strategies. For each group, we pretrain a specialized GNN query model using a sequential task scheduling strategy: tasks within the group are randomly shuffled and sequentially used to optimize the model over multiple episodes. By isolating tasks with compatible escalation strategies, we eliminate the conflicting updates of negative transfer, yielding specialized query models that are effective at uncovering PEs for specific untrusted entity types.

%% file: dataset.tex
\section{TAC-Bench}

\label{sec:taskgen}

Training TAC-GB's query models and evaluating TAC's detection capabilities require a large, diverse corpus of IAM PE tasks. The only public dataset, \texttt{IAM Vulnerable}~\cite{iamvulnerable}, contains just 31 simple tasks---too limited to represent enterprise-scale IAM configurations or support multi-task RL pretraining. To bridge this gap, we construct TAC-Bench, a benchmark of 2,500 LLM-synthesized and rigorously validated IAM PE tasks that capture the scale and diversity of real-world misconfigurations.

\begin{table*}[t]
	\centering
	\scalebox{0.9}{
	\begin{tabular}{|c|c|c|c|c|c|c|c|c|c|c|c|} 
		\hline
		\multirow{2}{*}{\textbf{Task Set}} & \multirow{2}{*}{\textbf{\# Task}} &
		\multicolumn{3}{c|}{\textbf{\# Entities}} & \multicolumn{3}{c|}{\textbf{\# Permissions}} & \multicolumn{3}{c|}{\textbf{\# Connections}} \\ 
		\cline{3-11}
		& & min & max & avg & min & max & avg & min & max & avg  \\ \hline
		\texttt{TAC-Bench} & 2,500 & 11 & 320 & 137 & 42 & 2,561 & 1,282 & 12 & 3,443 & 1,225  \\ \hline
		\texttt{IAM Vulnerable} & 31 & 2 & 3 & 2 & 1 & 6 & 2 & 3 & 12 & 6   \\ \hline
		\texttt{Startup} & 2 & 158 & 251 & 204 & 882 & 2,826 & 1,854 & 8,704 & 27,939 & 18,321 \\ \hline
	\end{tabular}
    }
	\caption{Statistics of three PE task sets.  }
		\label{tab:visible}
	\end{table*}

\noindent\textbf{IAM Misconfiguration Generation.}
Our goal is to automatically synthesize IAM configurations that mirror the size, richness, and complexity of real-world deployments, while ensuring each configuration contains a valid PE. To prevent the LLM from generating unrealistic or superficial vulnerabilities, we ground the generation process in authoritative sources---AWS official documentation~\cite{awsactiontable,awsbestpractice,awsiam}, prior systematic studies on IAM PEs~\cite{rhino:privesc,bishopstudy,sonraistudy,capitalonereport}, and all 31 tasks from \texttt{IAM Vulnerable}. Using OpenAI's gpt-5.4-mini\footnote{We employ OpenAI's \emph{gpt-5.4-mini}~\cite{gpt54mini} for its strong reasoning and structured planning capabilities, which are essential for generating, explaining, and validating PE chains. In addition, compared to full-sized reasoning models, gpt-5.4-mini delivers strong chain-of-thought performance at lower cost.}, we carry out a structured, three-stage generation process:

\begin{itemize} [nosep, left=0pt]
    \item \noindent\textbf{\emph{Stage~1: Realistic Context Generation.}}
We instruct the LLM to compose a realistic cloud organization description, grounded in public cloud deployment guidelines from cloud providers, government, and standards bodies~\cite{aws-well-architected,irs-cloud-computing-environment-bibtex,nist-sp-500-292-2011}. For diversity, we dynamically vary three dimensions across generations: business domain (e.g., healthcare, education, financial services), deployment scale (e.g., numbers of accounts, principals, and resources), and critical-asset profile (e.g., payment data, medical records, proprietary source code).

\item \noindent\textbf{\emph{Stage~2: Real-World PE Generation.}}
Given the organizational context, we instruct the LLM to instantiate an IAM misconfiguration that mimics real-world PE scenarios documented in the authoritative sources~\cite{awsactiontable,awsbestpractice,awsiam,rhino:privesc,bishopstudy,sonraistudy,capitalonereport}---including known attack techniques such as role chaining, policy attachment abuse, and credential manipulation. Few-shot examples from \texttt{IAM Vulnerable} are provided to illustrate the expected output format.

\item \noindent\textbf{\emph{Stage~3: Chain-of-Thought Verification.}}
Finally, the LLM must designate one \emph{untrusted entity} and one \emph{target permission}, and explicitly output the step-by-step propagation chain showing how the untrusted entity obtains the target permission through Chain-of-Thought reasoning. This both enforces the presence of a valid PE in the generated configuration and yields a human-readable explanation that we use for manual validation.
\end{itemize}

\noindent\textbf{PE Validation.}
Every generated configuration undergoes manual inspection. We independently verify each configuration's Chain-of-Thought reasoning step by step, confirming that every propagation step from the untrusted entity to the target permission corresponds to a valid IAM action with satisfied preconditions. This ensures that all 2,500 tasks in TAC-Bench contain genuine, validated PEs.

\noindent\textbf{Dataset Statistics.} 
Table~\ref{tab:visible} compares TAC-Bench with \texttt{IAM Vulnerable} and two real-world IAM misconfigurations we collected from a cloud security startup (\texttt{Startup}). We can observe that \texttt{IAM Vulnerable} configurations are extremely small (averaging 2 entities and 6 connections), making them unrepresentative of real-world cases. In contrast, TAC-Bench configurations average 137 entities and 1,282 permissions---closely matching the scale of two \texttt{Startup} configurations (avg.\ 204 entities and 1,854 permissions), which are drawn from real enterprise deployments. This demonstrates that TAC-Bench successfully bridges the gap between toy benchmarks and real-world complexity, providing realistic IAM PE tasks at scale.

%% file: eval.tex
\section{Evaluation}

\label{sec:eval}
\myparagraph{Pretraining and Evaluation Task Sets}
We partition the 2,500 tasks in TAC-Bench into two \emph{strictly disjoint} subsets: 2,000 tasks are used exclusively for multi-task pretraining of TAC-GB's query models, and the remaining 500 tasks are held out exclusively for testing. There is \emph{zero overlap}---no task used during multi-task pretraining ever appears in evaluation.

We evaluate TAC-WB and TAC-GB on three test sets spanning synthetic, public, and real-world settings:
(1)~\uline{\texttt{Test-Synthetic}}: the 500 held-out tasks from TAC-Bench described above;
(2)~\uline{\texttt{Test-Public}}: all 31 tasks from the only public benchmark \texttt{IAM Vulnerable}~\cite{iamvulnerable};
and (3)~\uline{\texttt{Test-Real}}: two large real-world IAM misconfigurations collected from a cloud security startup (the \texttt{Startup} task set in Table~\ref{tab:visible}). Although specific details of \texttt{Test-Real} cannot be disclosed due to data security protocols, both configurations contain at least one transitive PE with paths of five or more steps, posing significant detection challenges.
In total, 533 tasks are used for evaluation.

\myparagraph{Baselines}
For whitebox detection, we compare TAC-WB against three state-of-the-art publicly available PE detectors: \uline{\texttt{Pacu}}~\cite{pacu}, \uline{\texttt{Cloudsplaining}}~\cite{cloudsplaining}, and \uline{\texttt{PMapper}}~\cite{pmapper}. We exclude the approach by Shevrin and Margalit~\cite{shevrin2023detecting} as its artifact is not publicly available.

For greybox detection, no publicly available greybox or blackbox IAM PE detectors exist. We therefore construct four ablation variants of TAC-GB, each isolating a key design decision:
(1)~\uline{TAC-GB-RD}: selects queries randomly, measuring the benefit of our GNN-based RL approach;
(2)~\uline{TAC-GB-EA}: replaces RL with the evolutionary algorithm CMA-ES~\cite{hansen2006cma}, providing an alternative optimization strategy;
(3)~\uline{TAC-GB-NoPT}: removes pretraining entirely, isolating the contribution of offline multi-task pretraining;
(4)~\uline{TAC-GB-MamlPT}: replaces our entity-type-based pretraining with MAML~\cite{deleu2018effects}, evaluating how our pretraining strategy mitigates negative transfer.

\myparagraph{Evaluation Metrics} 
For effectiveness, we use \uline{false negative rate (FNR)} as the sole metric. All detectors---TAC-WB, TAC-GB, and the three baselines (\texttt{Pacu}, \texttt{Cloudsplaining}, \texttt{PMapper})---achieve 100\% precision by construction: every detector reports a PE \emph{only} when it produces a concrete escalation path in which each step is a documented IAM action whose preconditions are provably satisfied in the given configuration. Since precision is uniformly perfect, FNR is the only metric that differentiates detector effectiveness.

For efficiency, we measure \uline{runtime} for whitebox detectors and \uline{query count} for greybox detectors (i.e., the number of customer interactions required during detection).

Since greybox detection involves stochastic query selection and randomized customer responses, we report \uline{standard deviations} of both FNR and query count across 11 repeated experiments.  

\myparagraph{Greybox Detection Settings}

\noindent\textbf{\emph{Partial IAM configurations.}} For each task, we uniformly sample a subset of \emph{visible entities} from the full entity set and remove all permissions and relations involving non-visible entities, yielding a partial configuration.

\noindent\textbf{\emph{Customer query simulator with query declines.}} Since greybox detection requires interactive customer responses, we design a simulator that faithfully models both accepted and declined queries---directly reflecting the privacy-driven refusals central to the greybox setting (Section~\ref{subsec:greybox-pe-detection}). For each task, the simulator pre-samples a subset of queries from the query space that the customer is willing to answer. When a detector issues a query, the simulator checks membership in this set: if the query is in the set (\emph{accepted}), the simulator returns the ground-truth answer (\texttt{true} or \texttt{false}); if not (\emph{declined}), it returns \texttt{unknown}, and the query budget is still consumed. This means that under high rejection rates, TAC-GB must detect PEs with even fewer informative responses, directly increasing difficulty. We systematically study this effect by evaluating across 10 query budgets ($10$--$100$) for \texttt{Test-Synthetic} and \texttt{Test-Real}, where each budget represents at most 1.2\% of the 8,448-query space, and budgets of 10 and 20 for \texttt{Test-Public} (query space of 27). These tight budgets simulate realistic scenarios where customers answer only a small fraction of possible questions.

\noindent\textbf{\emph{Parameters.}} During pretraining, each task runs for 20 episodes. We use the AdamW optimizer~\cite{loshchilov2017decoupled} with a learning rate of \(10^{-4}\) for both pretraining and testing, and repeat all experiments 11 times.

%% file: experiments.tex
\begin{figure}[t!]
    \centering
    \includegraphics[width=\columnwidth]{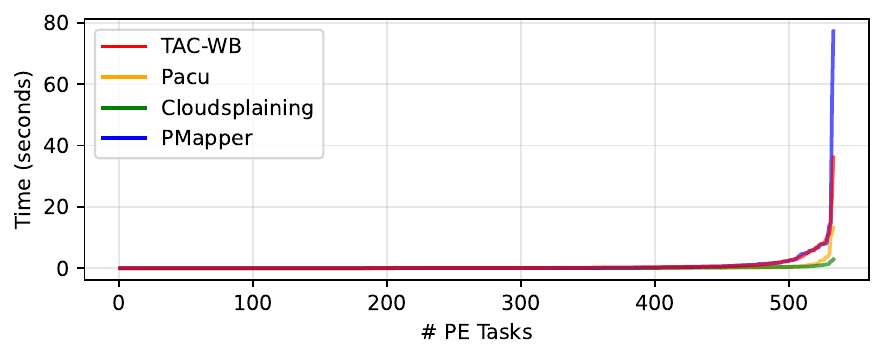}
    \caption{Time cost of whitebox detectors on all PE tasks.}
    \label{fig:eval:timecost}
\end{figure}

\begin{figure}[t]
	\centering
\begin{subfigure}[b]{\columnwidth}
			\centering
				\includegraphics[width=\columnwidth]{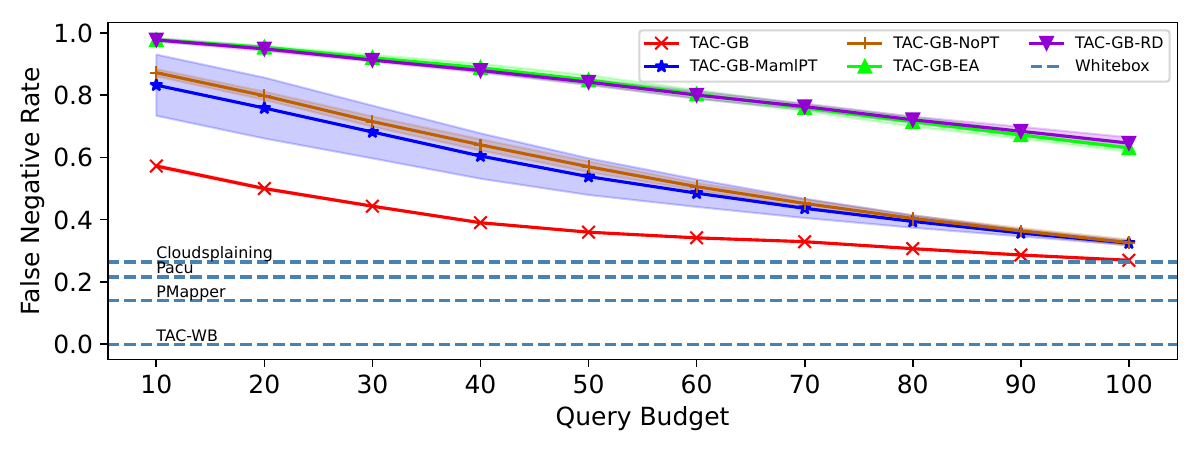}
	        \caption{False negative rates on \texttt{Test-Synthetic} across 10 query budgets.
				}
			\label{fig:eval:fnr}
\end{subfigure}
\vfill
		\begin{subfigure}[b]{\columnwidth}
				\centering
				\includegraphics[width=\columnwidth]{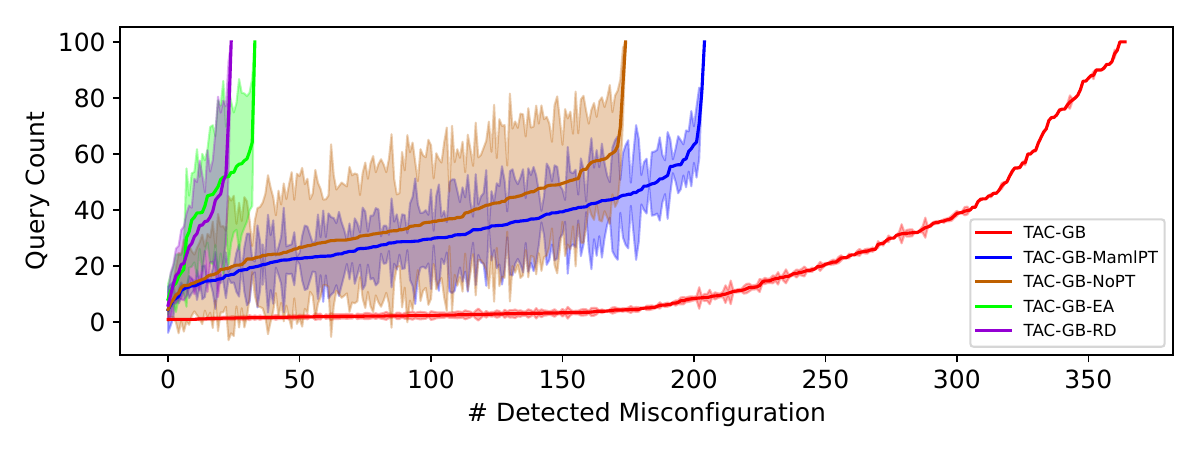}
			\caption{Query efficiency on \texttt{Test-Synthetic} under query budget 100. 
				}
				\label{fig:eval:aqr}
		\end{subfigure}
		\caption{Evaluation results on \texttt{Test-Synthetic}. Solid lines represent the mean, and shaded regions indicate the standard deviation across 11 repeated experiments.}
\end{figure}
\subsection{Experimental Results}
\label{sec:eval:exp}
We evaluate four research questions (RQ) to assess the effectiveness and efficiency of TAC-WB and TAC-GB, respectively.   
\subsubsection{RQ1: Effectiveness of TAC-WB}
TAC-WB detected all PEs across \texttt{Test-Synthetic}, \texttt{Test-Public}, and \texttt{Test-Real}, achieving a 0\% FNR. In contrast, \texttt{PMapper}, \texttt{Pacu}, and \texttt{Cloudsplaining} achieved FNRs of 12\%, 20\%, and 26\% on \texttt{Test-Synthetic}; 29\%, 32\%, and 39\% on \texttt{Test-Public}; and 100\% on \texttt{Test-Real} (all three failed to detect either of the two real-world PEs).

\noindent\textbf{Error analysis.} We manually inspected missed PEs by each baseline on \texttt{Test-Synthetic} (60 for \texttt{PMapper}, 100 for \texttt{Pacu}, 130 for \texttt{Cloudsplaining}) and classified them with our identified five PE types (Table~\ref{tab:permflow-categories}). 
\texttt{PMapper} models only user/role authentication chains, so it entirely misses three PE categories: Permission Delegation, Code Execution, and Credential Takeover. The pattern-based detectors encode only small, fixed catalogs of known attack patterns (\texttt{Pacu}: 21 patterns~\cite{rhino:privesc}; \texttt{Cloudsplaining}: a similarly limited set). Therefore, \texttt{Pacu} and  \texttt{Cloudsplaining} fail on tasks involving permission flows outside their predefined patterns, particularly in Permission Delegation and Code Execution.

On  \texttt{Test-Real}---which contains multi-step transitive PEs spanning five or more steps---all three baselines fail completely (100\% FNR), as none can reason about long propagation chains across multiple PE categories.

\noindent\textbf{Why TAC-WB achieves 0\% FNR.} The zero FNR follows directly from TAC-WB's exhaustive template coverage: 219 permission flow templates systematically extracted from the complete AWS operation set~\cite{awsactiontable} (14,000+ operations), spanning all five PE categories (Section~\ref{sec:study}).  Crucially, this result is confirmed across three independently sourced test sets: \texttt{Test-Synthetic} (LLM-generated), \texttt{Test-Public} (an external benchmark), and \texttt{Test-Real} (real enterprise configurations). This consistency across diverse, independently constructed datasets rules out overfitting to any single data source. Furthermore, across all 533 detected PE tasks, 206 out of 219 templates (94.1\%) are utilized, confirming that the broad template set is not redundant but actively contributes to detection.

\subsubsection{RQ2: Efficiency of TAC-WB}
Figure~\ref{fig:eval:timecost} reports the runtime of all whitebox detectors across \texttt{Test-Synthetic}, \texttt{Test-Public}, and \texttt{Test-Real}. For the vast majority of tasks, all detectors, including TAC-WB, complete within a second. On a small number of more complex tasks involving hundreds of permissions and flows, TAC-WB's runtime is comparable to or sometimes even faster than PMapper, the best-performing baseline, while still offering broader coverage. The faster runtimes of Pacu and Cloudsplaining stem from their pattern-based nature: they bypass multi-step (transitive) PEs and only match fixed patterns, which explains their lower computational cost but also their much higher false negative rates. \emph{In summary, TAC-WB achieves 0\% FNR while maintaining runtimes on par with the best-performing baseline, demonstrating that comprehensive PE coverage and practical efficiency are not in conflict.}
\subsubsection{RQ3: Effectiveness of TAC-GB}  
On \texttt{Test-Synthetic} (Figure~\ref{fig:eval:fnr}), TAC-GB consistently outperforms all four greybox variants across every query budget. Its FNR drops from 57\% to 27\% as the budget increases from 10 to 100, with near-zero standard deviation---demonstrating both the effectiveness and stability of our GNN-based RL approach with pretraining. At budget 100, the best variant TAC-GB-MamlPT still has a 59\% FNR, followed by TAC-GB-NoPT (65\%), TAC-GB-EA (93\%), and TAC-GB-RD (even higher), all with substantially larger standard deviations. Notably, even with only partial configuration access, TAC-GB's 27\% FNR closely rivals the whitebox baselines that have full access: \texttt{Cloudsplaining} (26\%), \texttt{Pacu} (20\%), and \texttt{PMapper} (12\%).

On \texttt{Test-Public}, TAC-GB detects 23 of 31 PEs at budget 10 (26\% FNR) and all 31 PEs at budget 20 (0\% FNR)---matching TAC-WB and outperforming every whitebox baseline (29\%--39\% FNRs). The greybox variants, by contrast, detect only 17--22 PEs at budget 10 (29\%--45\% FNRs) and 23--30 PEs at budget 20 (3\%--26\% FNRs).

On \texttt{Test-Real}, TAC-GB is the only detector to succeed, detecting both real-world PEs within a budget of 60, while all whitebox baselines (100\% FNR) and all greybox variants fail completely.

\emph{In summary, TAC-GB not only dominates all greybox variants but also rivals---and on \texttt{Test-Public} and \texttt{Test-Real} outperforms---whitebox baselines that require full configuration access, validating the practical viability of greybox PE detection.}
\subsubsection{RQ4: Efficiency of \Toolg}
Figure~\ref{fig:eval:aqr} compares the query counts of TAC-GB and its four greybox variants under a budget of 100 on \texttt{Test-Synthetic} (detected PEs for each detector are sorted by average query count across 11 repeated experiments).
TAC-GB detects 362 PEs with an average of only 19 queries per PE. The variants are both less effective and less efficient: TAC-GB-MamlPT detects 204 PEs with 1.7$\times$ higher average query count, TAC-GB-NoPT detects 175 PEs at 1.8$\times$, TAC-GB-EA detects just 34 PEs at 2.2$\times$, and TAC-GB-RD performs even worse. All four variants also exhibit substantially higher standard deviations, \emph{confirming that GNN-based RL with pretraining enables TAC-GB to be not only more query-efficient but also more stable across runs.}

%% file: discussion.tex
\section{Discussion}
\label{sec:discussion}

\myparagraph{Template Extensibility}
TAC's broad PE coverage (0\% FNR in our experiments) is tied to the 219 manually extracted permission flow templates. As AWS adds services, new permission flow types may emerge. The extraction methodology, however, is incrementally extensible: new templates are derived by reviewing newly added actions in the AWS actions table~\cite{awsactiontable} for permission propagation, and are incorporated without modifying TAC's detection algorithms.

\myparagraph{Scalability}
TAC-WB's fixed-point analysis runs in polynomial time and converges because the permission space is finite and monotonically non-decreasing; on the largest real-world configuration (251 entities, 2,826 permissions, 27,939 connections) in our \texttt{Test-Real} dataset, it completes within 40 seconds. TAC-GB's GNN query model (GATv2~\cite{brody2021attentive}) scales linearly with the number of edges and requires only a single forward pass per query---negligible compared to the human response time.
Crucially, PFG size grows with the number of entities and permission flows, not raw policy statements, because our permission space abstraction filters out PE-irrelevant permissions (Section~\ref{subsec:PFModeling}). For extremely large deployments (thousands of roles, millions of policies), partitioning by account or organizational unit could further bound graph size; we leave this as future work.

\myparagraph{Implications for Practitioners}
An internal security team can integrate TAC-WB into their pipeline to scan IAM configuration changes on every commit, catching PE-introducing misconfigurations \emph{before} deployment---analogous to running a static analysis linter, but for access control. For third-party auditors, TAC-GB opens a fundamentally new service model: rather than demanding a full configuration, auditors can conduct an interactive query session---our experiments show that 10--100 queries often suffice---while customers retain control over what they disclose, transforming PE detection from an all-or-nothing disclosure to a \emph{graduated, privacy-preserving engagement}.

In addition, although TAC currently targets AWS IAM, its core architecture is not AWS-specific. Any access control system where privileges propagate through hierarchical or delegated relationships can be modeled as a PFG by defining platform-specific templates. For example, Azure RBAC role assignments~\cite{azurerbac}, Kubernetes RBAC bindings~\cite{k8srbac}, and Google Cloud IAM policy inheritance~\cite{gcpiam} all exhibit propagation patterns analogous to those we capture. Adapting TAC to these platforms requires extracting new templates from their respective action spaces, while the detection algorithms and greybox query mechanism remain unchanged, making TAC a potential foundation for unified, cross-platform PE detection.

\myparagraph{Software Engineering Research Contributions}
The core challenge TAC-GB solves---detecting a property of a system whose specification is only partially observable, through adaptive interaction with a human stakeholder---arises in many SE contexts. Examples include privacy-preserving code review (where reviewers cannot see proprietary modules), third-party API conformance testing (where internal service logic is hidden), and compliance auditing of legacy systems with incomplete documentation. Our greybox approach provides a concrete, reusable blueprint for these settings.

TAC-WB's fixed-point analysis guarantees soundness, while TAC-GB's RL-guided queries optimize efficiency---but both operate over the same PFG representation. This shared abstraction allows formal guarantees and data-driven optimization to compose cleanly. This neuro-symbolic architecture---formal analysis for correctness, learned policies for efficiency---offers a promising design pattern for SE tools that require both soundness guarantees and practical scalability, such as automated program repair or program synthesis.

%% file: related_work.tex
\section{Related Work}

\label{sec:related}
All existing IAM PE detectors are whitebox and fall into three categories. \emph{Pattern-based} tools (\texttt{Pacu}~\cite{pacu}, \texttt{Cloudsplaining}~\cite{cloudsplaining}) encode small, fixed catalogs of  known PE patterns and generally miss transitive PEs~\cite{iamassess}. \emph{Graph-based} tools (\texttt{PMapper}~\cite{pmapper}, \texttt{AWSPX}~\cite{awspx}) model only user/role authentication chains, capturing a single PE category. Both families achieve only partial PE coverage (as shown in Table~\ref{tab:permflow-categories}), and our experiments confirm that these gaps lead to substantial missed PEs in practice.

The most closely related work is the \emph{reasoning-based} detector by Shevrin and Margalit~\cite{shevrin2023detecting}, which also reasons about how permissions change through AWS actions (their artifact is not publicly available, precluding direct experimental comparison). They encode IAM configurations as a finite-state Boolean model and apply bounded model checking to detect transitive PEs. However, the abstraction level differs fundamentally from TAC's permission flows. Shevrin and Margalit model individual AWS action semantics as Boolean transition relations---each action (e.g., \texttt{PutRolePolicy}, \texttt{AssumeRole}) requires hand-crafted logical formulas encoding its effect on the IAM state, and their coverage is tied to whichever actions the authors chose to formalize (primarily Rhino's 21 methods~\cite{rhino:privesc}). TAC abstracts at a higher level: permission flow templates group operations by \emph{source--sink entity-type pairs}, enabling systematic coverage of 14,000+ AWS actions through 219 templates rather than per-action Boolean encoding. This higher-level abstraction yields three concrete advantages: (1)~polynomial-time fixed-point analysis instead of bounded model checking (which faces state-space explosion), (2)~incremental extensibility---new templates are added without modifying the detection algorithm, and (3)~natural extension to greybox settings, which action-level model checking does not support.

%% file: conclusion.tex
\section{Conclusion} 
\label{sec:conclusion}
We presented TAC, the first hybrid framework unifying whitebox and greybox IAM privilege escalation detection. Built on the permission flow abstraction, TAC-WB achieves comprehensive PE coverage via 219 templates and fixed-point analysis, while TAC-GB---the first greybox PE detector---identifies escalations from partial configurations through RL-guided interactive queries. To support training and evaluation, we constructed TAC-Bench, a benchmark of 2,500 diverse IAM PE tasks that match the scale and complexity of real-world cases. Beyond IAM, TAC's neuro-symbolic architecture provides a reusable design pattern for SE tools that must balance soundness with adaptivity. In addition, its greybox formulation offers a practical blueprint for security analysis under partial observability, applicable to other cloud platforms and domains.